\documentclass[10pt,twocolumn,aps,pra,superscriptaddress,showpacs,tightenlines,pdflatex,longbibliography]{revtex4-2}
\usepackage{upgreek}
\usepackage{amsthm}
\usepackage{gensymb}
\usepackage{braket}
\usepackage{amsmath}            
\usepackage{amssymb}            
\usepackage{mathtools}
\usepackage{graphicx}           
\usepackage[caption=false]{subfig}
\usepackage{xcolor}

\usepackage{etoolbox}
\usepackage{breqn}

\usepackage{enumitem}
\usepackage{xcolor}
\usepackage{bm}
\usepackage{amsfonts}
\newcommand{\overbar}[1]{\mkern 1.5mu\overline{\mkern-1.5mu#1\mkern-1.5mu}\mkern 1.5mu}
\makeatletter
\let\cat@comma@active\@empty

\usepackage[colorlinks,citecolor=blue,urlcolor=blue,bookmarks=false,hypertexnames=true]{hyperref} 

\newcommand{\ts}{\otimes}
\newcommand{\id}{\openone}
\newcommand{\Tr}{\text{Tr}}

\newcommand{\ext}{\text{ext}}

\newcommand{\red}[1]{{\color{red} #1}}

\newcommand{\comment}[1]{}

\newtheorem{theorem}{Result}

\begin{document}
\title{A General Class of Functionals for Certifying Quantum Incompatibility}

\author{Kuan-Yi Lee}
\affiliation{Scuola Normale Superiore, Piazza dei Cavalieri 7, 56126 Pisa, Italy}
\affiliation{Department of Physics and Center for Quantum Frontiers of Research \&
Technology (QFort), National Cheng Kung University, Tainan 701, Taiwan}

\author{Jhen-Dong Lin}
\email{jhendonglin@gmail.com}
\affiliation{Department of Physics and Center for Quantum Frontiers of Research \&
Technology (QFort), National Cheng Kung University, Tainan 701, Taiwan}

\author{Adam Miranowicz}
\affiliation{Institute of Spintronics and Quantum Information, Faculty of Physics and Astronomy, Adam Mickiewicz University, 61-614 Pozna\'n, Poland}

\author{Yueh-Nan Chen}
\email{yuehnan@mail.ncku.edu.tw}
\affiliation{Department of Physics and Center for Quantum Frontiers of Research \& Technology (QFort), National Cheng Kung University, Tainan 701, Taiwan}
\affiliation{Physics Division, National Center for Theoretical Sciences, Taipei 106319, Taiwan}

\date\today

\begin{abstract}
Quantum steering, measurement incompatibility, and instrument incompatibility have recently been recognized as unified manifestations of quantum incompatibility. 
Building on this perspective, we develop a general framework for constructing optimization-free, nonlinear incompatibility witnesses based on convex functionals, valid in arbitrary dimensions. 
We prove that these witnesses are nontrivial precisely when the underlying functional is non-affine on extremal points (e.g., pure states for ensembles). 
For pure bipartite states, the witnesses yield lower bounds on entanglement measures, thereby outperforming most linear steering inequalities in the pure-state regime.
Moreover, the construction extends in full generality to certify measurement and instrument incompatibility, where the witnesses act as genuine incompatibility monotones. 
We demonstrate the versatility of our approach with two operationally relevant functionals: the Wigner-Yanase skew information and an $\ell_{2}$-type coherence functional.

\end{abstract}

\maketitle

\textit{Introduction.}---Incompatibility is a foundational quantum resource.
In classical physics, any prescribed family of operations can be reproduced by a single device together with classical post-processing~\cite{Heinosaari2016,Guhne2023rmp}.
By contrast, many quantum families admit no such joint simulation, i.e., compatibility fails.
This structural constraint, already exemplified by the position–momentum trade-off, powers steering~\cite{Gallego2015prx,Xiang2022prx,Uola2023RMP}, Bell nonlocality~\cite{Buscemi2012prl,Brunner2014rmp}, cryptography~\cite{Gisin2002rmp,Branciard2012pra}, and precision advantages in estimation and discrimination~\cite{Piani2015prl,Takagi2019prl,Skrzypczyk2019prl,Paul2019prl,Tan2021prl,Flatt2022prx,Lee2023prr}.

An increasing number of studies emphasize a unified compatibility viewpoint for states, measurements, and instruments~\cite{Uola2018pra,Guhne2023rmp}. 
In this formulation, quantum steering is the failure of a local-hidden-state model~\cite{Wiseman2007PRL} (the compatibility of state assemblage), its link to measurement incompatibility ties (un)steerability to (joint) measurability~\cite{Quintino2014PRL,Uola2014PRL,Uola2015PRL,Ku2022NC}.
Related developments extend the same framework to temporal and channel scenarios~\cite{Chen2014pra,Bartkiewicz2016pra,chen2016,Buscemi2020prl,Leppjrvi2024,Ji2024prx}, allowing assemblages across settings to be analyzed on comparable footing.

For certification, two approaches dominate: first, optimization-based measures (e.g., robustness and compatibility via semi-definite programs) are informative but computationally demanding~\cite{Cavalcanti_2016,Designolle2019}; and, second, linear or non-linear witnesses (e.g., moment-matrix criteria~\cite{Kogias2015prl,Chen2016prl}, uncertainty-based inequalities~\cite{Pramanik2014pra,Krivachy2018pra,Costa2018pra}, and tests based on the Clauser-Horne-Shimony-Holt inequality~\cite{Wolf2009prl,Cavalcanti2015}) are easy to implement yet sometimes weak or lacking a clear task-related interpretation.
This gap motivates simple, model-independent tests that avoid optimization while retaining operational meaning.

In this work, we develop a convex-analytic framework to assemblages and show that \emph{any} convex functional yields a class of optimization-free, non-linear, and dimension-independent incompatibility witnesses. 
We prove a necessary and sufficient criterion: a witness of quantum incompatibility is nontrivial if and only if the functional is non-affine on extremal points.
The framework delivers two concrete, operationally meaningful witnesses--instantiated with the Wigner–Yanase skew information (WYSI)~\cite{Wigner1963,Luo2003prl} and an $\ell_2$-type coherence functional~\cite{Baumgratz2014PRL}--and encompasses earlier metrology-~\cite{Yadin2021NC} and coherence-based~\cite{Lee2025npj} tasks within a single framework. 

For pure states, the witness lower-bounds an entanglement measure and is saturated only by maximally entangled states. 
Via the steering–equivalent-observable map and minimal Stinespring dilation~\cite{Uola2018pra}, the construction applies uniformly to measurements and instruments, recovering the known joint-measurability thresholds~\cite{Heinosaari2008} and illustrating distinctive channel behavior.

\begin{figure}
    \centering
    \includegraphics[width=0.99\linewidth]{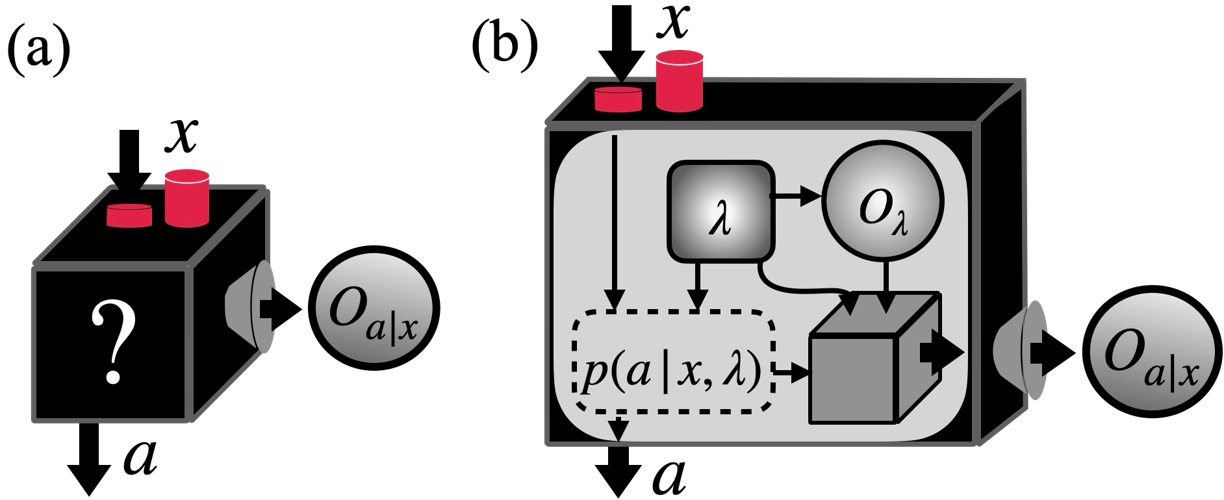}
    \caption{Schematic illustration of compatibility tasks.}
    \label{fig1}
\end{figure}

\textit{A unified framework of quantum incompatibility test.}---It is known that quantum steering, measurement incompatibility, and instrument incompatibility can be investigated within a single unified framework in semi-device-independent settings~\cite{Uola2018pra,Uola2023RMP,Guhne2023rmp}. 
We refer to this framework as quantum incompatibility. 
Specifically, as depicted in Fig.~\ref{fig1}\red{(a)}, we consider a quantum device with multiple classical inputs labeled by $x$, each corresponding to a button. 
When a particular input is chosen, the device stochastically produces a classical output $a$ with probability $p(a|x)$, along with a corresponding quantum object $O_{a|x}$. 
We note that this quantum object can generally represent a quantum state, a quantum measurement, or a quantum instrument with corresponding notations in Tab.~\ref{Table}.
We temporally leave it unspecified to maintain the generality. 

\begin{table}[!hbpt]
\begin{tabular}{c|c|c}
     Spaces         & ~~Assemblages ~~             &  ~~Witnesses ~~ \\\hline
 States  & $\boldsymbol{\sigma}$   & $\mathcal S_{g}(\boldsymbol{\sigma})$ \\
 ~~Measurements~~  & $\boldsymbol{M}$        & $\mathcal M_{g}(\boldsymbol{M})$ \\
 ~~Instruments~~ & $\boldsymbol{\Lambda}$  & $\mathcal I_{g}(\boldsymbol{\Lambda})$
\end{tabular}
\caption{Notations of assemblages and witnesses.}\label{Table}
\end{table}

This device can now be fully characterized by an assemblage $\boldsymbol{O}=\{p(a|x)O_{a|x}\}_{a,x}$, which allows us to introduce the notion of quantum incompatibility. 
We regroup the assemblage according to each input $x$, defining $\boldsymbol{O}_x=\{p(a|x)O_{a|x}\}_{a}$, and ask whether $\boldsymbol{O}_x$ are mutually compatible. 
If this is the case, the assemblage can be classically simulated by a hidden-object (HO) model as depicted in Fig.~\ref{fig1}\red{(b)}. 
This model consists of predetermined hidden parent objects $\{O_\lambda\}_\lambda$ drawn from a distribution $p_\lambda$ and a classical stochastic map $p(a|x,\lambda)$ such that  
\begin{equation}
\begin{aligned}
    p(a|x)O_{a|x} = \sum_\lambda p(\lambda)p(a|x,\lambda)O_\lambda.\label{eq: HOmodel}
\end{aligned}
\end{equation}
An assemblage is quantum incompatible, if it cannot be simulated by the HO model. 

We are now in position to present our main result, in which we propose a framework comprising a general class of information tasks to certify quantum incompatibility. 
The central ingredient of this approach relies on the concept of real-valued convex functionals of quantum objects. 
Specifically, let $\mathcal O$ be the space of all quantum objects. A convex functional $g :\mathcal{O} \to \mathbb{R}^{+}$ satisfies $\sum_i p_i g(O_i) \ge g(\sum_i p_i O_i)$ with $p_i\geq 0$ and $\sum_i p_i=1$. 
Operationally, most resource quantifiers satisfy convexity, which implies that classical randomness and mixture cannot create non-classical resources.
For every setting $x$, we define a functional of $\boldsymbol{O}_x$ as
\begin{equation}
    g^{\rm as}(\boldsymbol{O}_x):= \sum_a p(a|x)g(O_{a|x}),\label{eq: gas}
\end{equation}
where superscript ``as'' refers to a functional of assemblage.
In addition, we define the roof extension of $g$ as $F_{g}(O):=\sup_{ \{ p_i,\psi_i\}_i } \sum_i p_i g(\psi_i)$ that satisfies $\sum_{i} p_i \psi_i = O$ and $\{\psi_i\}_{i} \subset \ext(\mathcal{O})$, where $\rm ext (\mathcal{O})$ denotes the set of extreme points of $\mathcal{O}$, e.g., pure states for a state space.
Note that the roof extension is achievable provided the supremum is attained by a physically realizable decomposition.
Due to the convexity of $g$ in $\mathcal{O}$, the roof extension serves as an upper bound of $g$. Similarly, we define $F^{\rm as}_{g}(\boldsymbol O_x):=\sum_a p(a|x) F_{g}(O_{a|x})$. Following the standard convex-theoretic analysis, we obtain the following general criterion for certifying quantum incompatibility. 
\begin{theorem}
   If an assemblage $\boldsymbol{O}$ admits the HO model, then
\begin{equation}
    g^{\rm as}(\boldsymbol{O}_{x})\leq F_{g}^{\rm as}(\boldsymbol{O}_{x'})~~\forall x,x^{\prime}.\label{eq: HO ineq}
\end{equation} 
\end{theorem}
The details of our proof can be found in the Appendix~\ref{appendix: HO ineq}, where the central idea relies on the convexity (concavity) of $g$ ($F_g$) by showing that $g^{\rm as}(\boldsymbol{O}_x)\le \sum_\lambda p(\lambda) g(O_\lambda)$ $[F^{\rm as}_g(\boldsymbol{O}_x)\ge \sum_\lambda p(\lambda)F_g(O_\lambda)]$, respectively.

If the HO model holds, each conditional object $O_{a|x}$ is generated from the same parent ensemble $\{ O_\lambda \}_\lambda$ by classical data processing. 
Hence, the observed average over any setting $g^{\rm as}(\boldsymbol{O}_x)$ cannot exceed the value obtainable from $\sum_\lambda p(\lambda)g(O_\lambda)$.
On the other hand, $F_g^{\mathrm{as}}(\boldsymbol{O}_{x})$ is the maximal average value of $g$ achievable from the same $\{ O_\lambda \}_\lambda$ when one has full classical information about the hidden variable $\lambda$ and is free to optimally from the outcome labels $(a,x)$.
As such, Eq.~\eqref{eq: HO ineq} states that, when the assemblage can be generated by data‑processing of a single parent object $\{O_{\lambda}\}_\lambda$, the resource observed in a specific task can never surpass the information-assisted formation limit based on the same ingredients.
A violation therefore certifies the impossibility of such data‑processing, and its amount indicates the extra resource enabled by the incompatibility.
We note that an entirely parallel construction holds for any concave function $f$ and its convex roof $G_{f}$ such that $f\geq G_{f}$.
For conciseness, in the following, we only discuss the convex function $g$.

One can ask two natural questions:
First, can the inequality be violated at all, say, the inequality is non-trivial?
Second, how can one determine from $g$ whether the resulting inequality is trivial or non-trivial? 
To address these questions, we establish the necessary and sufficient condition for Eq.~\eqref{eq: HO ineq} to be non-trivial and provide explicit examples illustrating both the trivial and non-trivial cases.
For convenience, we define the violation degree of Eq.~\eqref{eq: HO ineq} as a witness:
\begin{equation}
    \mathcal{V}_{g}(\boldsymbol{O}):= \max \left\{ \max_{x} g^{\rm as}(\boldsymbol{O}_{x}) - \min_{x} F_{g}^{\rm as}(\boldsymbol{O}_{x}), 0 \right\},\label{eq: violation}
\end{equation}
which vanishes for every HO assemblage by Result 1.

Below we give a necessary and sufficient condition for $\mathcal V_{g}$ to be non-trivial--that is, to attain strictly positive values. 
Trivial choices of $g$ follow immediately from this statement, followed by two non-trivial examples.

\begin{theorem}
    $\mathcal{V}_{g}$ is non-trivial, i.e., $\exists\,\boldsymbol O^{\prime}$ such that $\mathcal V_{g}(\boldsymbol O^{\prime})>0$, if and only if $g[\text{ext}(\mathcal{O})]$ is not affine.
\end{theorem}

The details are presented in 
the Appendix~\ref{appendix: nontrivial}.
An affine functional is simultaneously convex and concave, hence invariant under both data processing and convex decomposition. 
Consequently, Eq.~\eqref{eq: HO ineq} holds with equality if $g$ is affine on the whole $\mathcal O$, rendering the criterion trivial.
However, non-affineness on $\mathcal O$ alone is not sufficient to being non-trivial: if the restriction $g|_{\mathrm{ext}(\mathcal O)}$ is affine (e.g., constant along extremal segments), the inequality remains trivial. 
An example is ergotropy~\cite{Alicki2013rpe} defined as
\begin{equation}
    \mathcal{E}(\rho) := \max_{U} \Tr\, H(\rho - U\rho \, U^{\dagger}),\label{eq: ergo} 
\end{equation}
where $H$ is a fixed system Hamiltonian and $U$ ranges over all unitaries. 
Ergotropy is convex on the state space, yet for any \emph{pure} input state Eq.~\eqref{eq: ergo} reduces to an affine form $F_{\mathcal E}(\rho)=\Tr \rho H - h_{\min}$, where $h_{\min}$ is the smallest eigenvalue of $H$, so that $F_{\mathcal E}=F_{\mathcal E}^{\rm as}$ uniformly upper-bounds every $\mathcal{E}^{\rm as}$ [c.f. Eq.~\eqref{eq: gas}].
Consequently, the HO inequality collapses to an identity and becomes trivial.

Here, we provide two convex functionals as examples of constructing non-trivial inequalities.
The first one is the WYSI $I(\rho,H)=- \frac{1}{2}\Tr\,[\sqrt{\rho},H]^{2}$, which serves as a measure of quantum uncertainty associated with an observable $H$. Unlike the total variance, WYSI excludes contributions arising from classical mixing of the state~\cite{Girolami2013prl,Girolami2014prl}. 
In addition, when $H$ is interpreted as the generator of time evolution, the WYSI quantifies the asymmetry of the state $\rho$, bounding the intrinsic speed of quantum evolution~\cite{Marvian2016pra,Pires2016prx,Takagi2019sr,Yamaguchi2023prl}.
Note that the task of such an incompatibility scenario can be described as follows: 
After Bob receives the state assemblage, he can either measure the asymmetry by evaluating the WYSI on each conditional state $\rho_{a|x}$ and averaging, obtaining $I^{\rm as}$ by Eq.~\eqref{eq: gas}, or estimate an upper bound using only classical post-processing via the concave-roof envelope $F_{I}^{\rm as}$.
In such a scenario, if Bob discovers that there exists a pair $(x,x')$ such that the measured asymmetry surpasses the estimated upper bound in Eq.~\eqref{eq: HO ineq}, the assemblage must be incompatible.
Similar scenarios can also be applied for quantum Fisher information and distillable coherence (c.f. Refs.~\cite{Yadin2021NC,Lee2025npj}).
A systemic way of constructing the witness by $I(\rho,H)$ is given in the Appendix~\ref{appendix: examples}, followed by the second example: $\ell_2$-type of coherence $N(\rho,\boldsymbol{\Pi}):= \sum_{i\neq j} |\bra{i} \rho \ket{j}|^2$, which measures the off-diagonal quantum coherence within the fixed reference basis $\boldsymbol{\Pi}:=\{ \ket{i}\bra{i}\}_{i}$~\cite{Baumgratz2014PRL,Streltsov2017rmp}.
This quantity bounds the energy storage in quantum batteries~\cite{Caravelli2021} and detects localization transitions~\cite{Styliaris2019prb}.

\textit{State assemblage and steering scenario.}---In this section we apply our convex–analytic framework for the case where the object space $\mathcal{O}$ is the set of quantum states.
Under this identification, the task of assemblage incompatibility becomes the familiar task of Einstein-Podolsky-Rosen (EPR) steering~\cite{Gallego2015prx,Uola2023RMP}. 
We begin by recalling the standard steering protocol and then show how its conditional states form a state assemblage to which our criterion directly applies.

In a typical EPR steering experiment, Alice (A) and Bob (B) share a bipartite state $\rho^{\rm AB}$.
In each round Alice chooses a measurement setting $x$ and performs the positive-operator valued measure $\boldsymbol{M}_x = \{M_{a|x}\}_a$ with outcomes $a$, where $M_{a|x}\ge 0$ and $\sum_a M_{a|x}=\id^{\rm A}$.
After obtaining an outcome $a$, she sends the pair $(a,x)$ to Bob.
Bob's sub-normalized conditional state is $\sigma_{a|x}:=\Tr_{\rm A}[(M_{a|x}\otimes\id^B)\,\rho^{\rm AB}]$.
The collection $\boldsymbol{\sigma}:=\{\sigma_{a|x}=p(a|x)\rho_{a|x}\}_{a,x}$ is called a state assemblage satisfying
$\sum_a p(a|x)\rho_{a|x}=\rho^{\rm B}=\Tr_{\rm A}\,\rho^{\rm AB}$ for all $x$.
An assemblage is said to be \emph{unsteerable} if it admits a local-hidden-state (LHS) model~\cite{Wiseman2007PRL},
\begin{equation}
\begin{aligned}
    p(a|x)\rho_{a|x} 
    =& \sum_{\lambda}p(\lambda)p(a|x,\lambda)\rho_{\lambda},\\
\end{aligned}\label{eq: LHS model}
\end{equation}
where $\{\rho_\lambda\}_\lambda$ are pre-existing states on Bob’s side and $p(a|x,\lambda)$ is a classical response function.
Otherwise the assemblage (and hence the underlying bipartite state) is \emph{steerable}.
Consequently, given any functional of state that is convex and non-affine under pure states, i.e., extreme points of the state space, Eq.~\eqref{eq: violation} becomes a non-trivial steering witness; we denote the violation degree as the witness $\mathcal S_g (\boldsymbol{\sigma})$.

Moreover, one can notice that the witness $\mathcal{S}_g$ is not only determined by the underlying correlations of the shared state $\rho^{\rm AB}$, but it also depends on Alice's measurement $M_{a|x}$. 
In fact, inappropriate choices on Alice's side may hide non-classical correlations altogether, yielding little or no violation even when $\rho^{\rm AB}$ is highly non-classical~\cite{Quintino2014PRL,Uola2014PRL}. 
In the following, we temporarily set aside the problem of measurement design (which is analyzed in the next Section) and focus on how much violation degree can be attributed to the state itself.

To factor out the measurement dependence, we henceforth optimize over Alice’s measurements. 
For pure $\rho^{\rm AB}$, every ensemble of $\rho^{\rm B}$ is pure by performing rank-1 projective measurements by Alice~\cite{Hughston1993}, so the supremum of $g$ coincides with the concave-roof value.
In such a case, the steering violation achieves the maximum value without measurement dependency (see Appendix~\ref{appendix: entanglement measure} for the detailed proof) and possesses the following property:
\begin{theorem}
    If $g$ is convex and symmetric, $\mathcal{S}_{g}$ lower bounds a full entanglement measure when evaluating pure bipartite states, and the maximum violation degree $\mathcal{S}_g^{\rm max}=[F_{g}-g](\openone_{d})$ is achieved if only if the shared states $\rho^{\rm AB}$ is maximally entangled.
\end{theorem}

We prove this property by showing that when $\rho^{\rm AB}$ is pure entangled, $\mathcal{S}_{g}$ is upper bounded by a closed form $[F_g-g](\rho^{\rm B})$, which is concave and symmetric~\cite{Horodecki2009rmp}; the maximum violation is directly obtained by Schur-concavity.
Notably, the bound is saturated when the roof extensions are achievable, i.e., the optimal decompositions always exist. 
In such case, the $\mathcal{S}_g$ becomes a faithful entanglement measure for pure bipartite states.
A detailed proof can be found in the Appendix~\ref{appendix: entanglement measure}.
We remark that Result 3 shows that the algebraic maximum of $\mathcal{S}_g^{\rm max}$ is achieved only by maximally entangled states.
This certifies the underlying states up to local isometries, but, since our figure of merit is optimized over Alice’s measurements, it does not fix a unique measurement realization. 
From the perspective of self-testing~\cite{supic2020quantum,Sarkar2021npjqi}, this corresponds to a \textit{weak form} of self-testing, where a state is identified but the measurements remain unidentified~\cite{Kaniewski2020prr}.

\textit{Measurement assemblage and steering-induced incompatibility.}---Here, we restrict the object space to measurements with the corresponding assemblage $\boldsymbol{M}:=\{M_{a|x}\}_{a,x}$.
An assemblage of measurements is said to be \emph{incompatible} when the corresponding observables cannot be implemented simultaneously--an intrinsically quantum feature that underpins contextuality~\cite{Tavakoli2020prr,Budroni2022rmp} and uncertainty relations~\cite{Heinosaari2016,Mao2023prl}, and serves as a necessary resource for demonstrate EPR steering~\cite{Uola2023RMP}.
Specifically, a measurement assemblage $\boldsymbol{M}$ is \emph{compatible} if there exists a single measurement $G_{\lambda}\ge 0$ such
that
\begin{equation}
    M_{a|x} = \sum_{\lambda} p(a|x,\lambda)G_{\lambda} \quad \forall a,x,\label{eq: joint measurement}
\end{equation}
where $p(a|x,\lambda)$ is a conditional probability distribution.
Operationally, Eq.~\eqref{eq: joint measurement} means that the statistics generated by $\boldsymbol{M}$ can be simulated by performing a single measurement $\{G_\lambda\}_{\lambda}$ followed by post-processing.
Following Result 1, we can conclude that: For any convex $g$ of a measurement $M$, we can construct a witness of measurement incompatibility, denoted as $\mathcal{M}_g(\boldsymbol{M})$.
To adopt the same functionals, i.e., $I(\rho,H)$ and $N(\rho,\boldsymbol{\Pi})$, we demonstrate below a method for constructing a valid incompatibility measure through convex functionals on the state space.

The structure of Eq.~\eqref{eq: joint measurement} mirrors the LHS model in Eq.~\eqref{eq: LHS model}. 
Indeed, Refs.~\cite{Uola2014PRL,Quintino2014PRL,Uola2015PRL} establish a one‑to‑one correspondence between the measurement assemblage on Alice’s side and the state assemblage on Bob's side, testing incompatibility structures in the steering scenario by applying pure entangled states~\cite{Quintino2019prl}.
Specifically, Alice performs $\boldsymbol{M}$ and introduces state assemblage $\boldsymbol{\sigma}$ on Bob, one can define the steering-equivalent observables (SEO): $\boldsymbol{B}:=\{B_{a|x}=\rho^{\text{B}-1/2}\,\sigma_{a|x}\,\rho^{\text{B}-1/2}\}_{a,x}$.
Then $\boldsymbol{M}$ is compatible if and only if the corresponding $\boldsymbol{\sigma}$ is unsteerable (equivalently, if and only if $\boldsymbol{B}$ is compatible).
By doing so, one can map the steering witness $\mathcal{S}_{g}(\boldsymbol{\sigma})$ to a measurement‑incompatibility witness by setting $\boldsymbol{\sigma}\to \sqrt{\rho^{\rm B}}\,\boldsymbol{B}\sqrt{\rho^{\rm B}}=\sqrt{\rho^{\rm B}}\,\boldsymbol{M}\sqrt{\rho^{\rm B}}$.

In addition, to eliminate the explicit dependence on the choice of state, we consider the scenario of a steering-induced incompatibility measure~\cite{Ku2022NC,Hsieh2023} by optimizing $\rho^{\rm B}$, and we have
\begin{equation}
   \mathcal{M}_g(\boldsymbol{M}) = \sup_{\rho^{\rm B}} \mathcal{S}_g \left(\sqrt{\rho^{\rm B}}\,\boldsymbol{M}\sqrt{\rho^{\rm B}}\right), \label{eq: violation of measurement}
\end{equation}
where the supremum is taking over all full-rank $\rho^{\rm B}$~\cite{Uola2015PRL}.
In other words, given a convex $g:\mathcal{D}(\mathcal H)\to \mathbb{R}^+$, we can extend it into a functional that maps $\mathcal{L}(\mathcal H)\to \mathbb{R}^+$.
Furthermore, we show that $\mathcal M_g(\boldsymbol{M})$ serves as a valid incompatibility monotone~\cite{Paul2019prl}, as presented in the following result:
\begin{theorem}
    The steering-induced incompatibility $\mathcal{M}_{g}(\boldsymbol{M})$ quantifies incompatibility of $\boldsymbol{M}$ on Alice's side.
\end{theorem}
We prove this result by demonstrating that $\mathcal{M}_{g}(\boldsymbol{M})$ satisfies: (i) $\mathcal{M}_{g}(\boldsymbol{M})=0$ if $\boldsymbol{M}$ is compatible, (ii) $\mathcal{M}_{g}(\boldsymbol{M})$ is convex, and (iii) $\mathcal{M}_{g}(\boldsymbol{M})$ is non-increasing under post-processing.
The details are given in 
the Appendix~\ref{appendix: measurement assemblage}.

\begin{figure}[!hbpt]
\centering
\includegraphics[width=0.99\columnwidth]{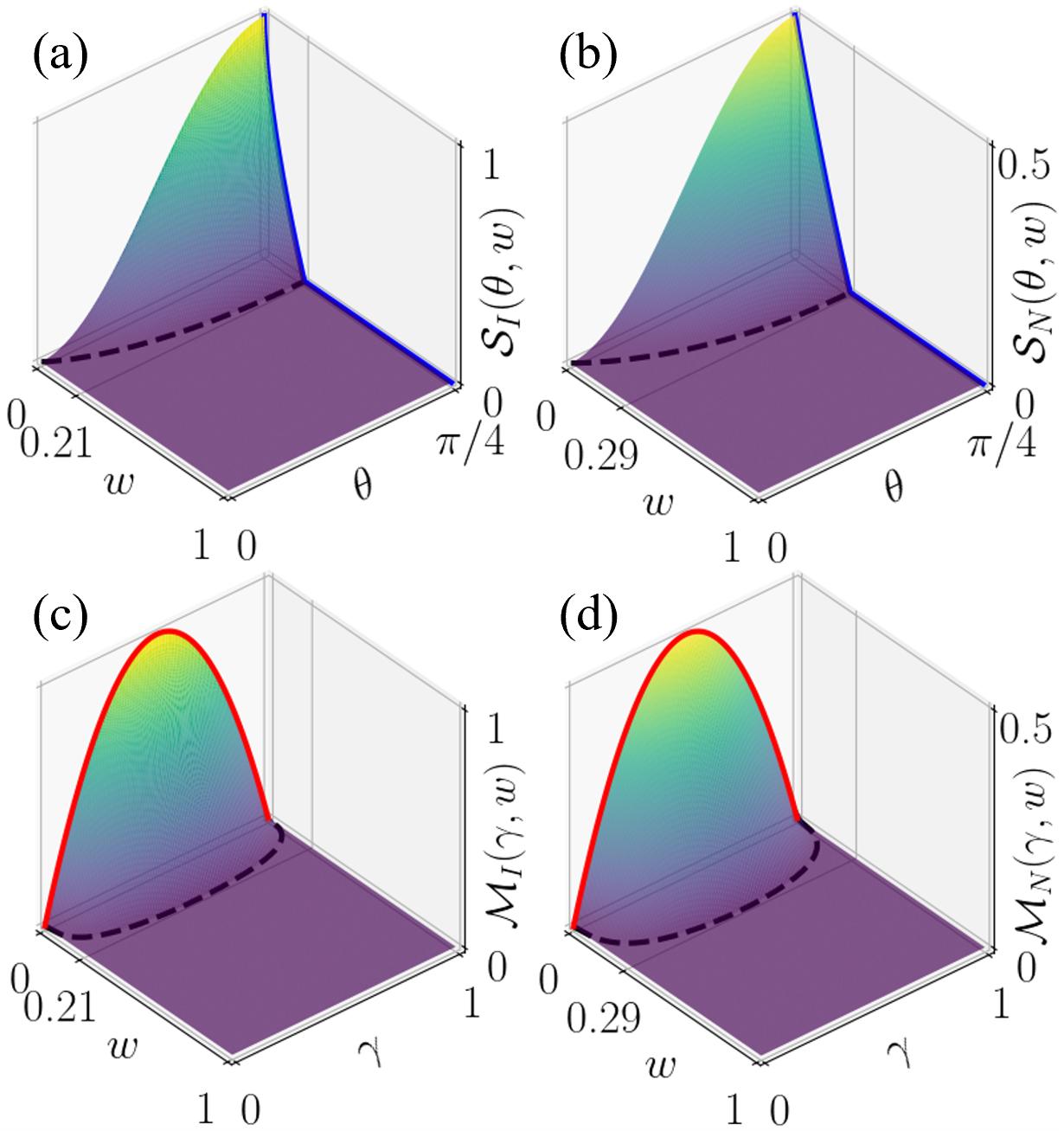}
\caption{
The violation degrees of (a) $\mathcal{S}_{I}(\theta, w)$ and (b) $\mathcal{S}_{N}(\theta, w)$, where the black-dashed curves represent the boundaries of zero violation.
The largest violations appear at $\theta = \pi /4$ are marked as blue curves; at these slices, $\mathcal{S}_I|_{\theta=\pi/4}=\mathcal M_I(w)$ and $\mathcal{S}_N|_{\theta=\pi/4}=\mathcal M_N(w)$ and both monotonically decrease when $w$ increases.
One can observe that the $\mathcal{M}_I$ is more sensitive to noise and vanishes as $w\approx 0.21$, while the $\mathcal M_N$ meets the threshold of compatible measurement, i.e., $w=1- 1/\sqrt{2}\approx 0.29$.
The violation degrees of (c) $\mathcal{M}_I(\gamma,w)$ and (d) $\mathcal{M}_N(\gamma,w)$.
Analogously, when $w=0$, $\mathcal{M}_I|_{w=0}$ and $\mathcal{M}_N|_{w=0}$ reduce to $\mathcal{I}_I(\gamma)$ and $\mathcal{I}_N(\gamma)$, respectively, indicating the instrument incompatibility (see the red curves in each figures).
Additionally, let $\gamma = 0.5$, the minimum dilation maps the input state $\ket{1}$ to a maximally entangled state, reflecting the measurement incompatibility $\mathcal M_I(\boldsymbol{M})$ and $\mathcal M_N(\boldsymbol{M})$.
}\label{fig2}
\end{figure}

To demonstrate our result, we consider a set of two noisy qubit Pauli measurements defined as
\begin{equation}
    M_{a|x}^w = (1-w)\Pi_{a|x} + w \frac{\id}{2},\label{eq: noisy Pauli}
\end{equation}
where $x \in \{0,1\}$ represents Pauli Z and X measurement with its corresponding eigen-projectors $\Pi_{a|x}$.
These measurements are incompatible for the mixing parameter $w\le 1- 1/\sqrt{2} \approx 0.29$~\cite{Heinosaari2008}.
We plot the value of $\mathcal{S}_{g}(\sqrt{\rho^{\rm B}}\,\boldsymbol{M}\sqrt{\rho^{\rm B}})$ with the pure entangled input $\ket{\varphi^{AB}(\theta)}=\sin \theta\ket{00}+\cos \theta\ket{11}$ in Figs.~\ref{fig2}\red{(a)} and \ref{fig2}\red{(b)}.
The black-dashed curves represent the boundary of zero violation degrees.
Taking $\theta=\pi /4$ (corresponding to optimizing $\rho^{\rm B}$ and marking as the blue curves), we obtain the largest violation degree for both examples.
For such slices, $\mathcal{S}_I (w)|_{\theta=\pi /4}$ and $\mathcal{S}_N|_{\theta=\pi /4}$ reduce to $\mathcal{M}_I (w)$ and $\mathcal{M}_N (w)$.
One can observe that, while the $\mathcal{M}_I (w)$ is more sensitive to noise (vanishes when $w\approx 0.21$), the $\mathcal{M}_N (w)$ is a tighter quantifier as it meets the threshold of $w\approx 0.29$ (the analytical results are shown in the Appendix.~\ref{appendix: measurement assemblage example}).

\textit{Instruments assemblage and channel steering.}---Now, we consider the object assemblage to instrument assemblage $\boldsymbol{\Lambda}=\{\Lambda_{a|x}\}_{a,x}$, where each $\Lambda_{a|x}$ is a completely-positive trace-non-increasing map.
For a fixed setting $x$, the family $\boldsymbol{\Lambda}_x=\{\Lambda_{a|x}\}_{a}$ constitutes an instrument satisfying the same marginal channel $\Lambda^{\rm m}=\sum_{a}\Lambda_{a|x}~\forall x$.
Physically, each element in an instrument $\{\Lambda_{a|x}\}_a$ acts as a general quantum measurement: it first post-selects the input state $\rho$ with a success probability $\Tr\,\Lambda_{a|x}(\rho)$, and then outputs the post-measurement state $ \Lambda_{a|x}(\rho)/\Tr\,\Lambda_{a|x}(\rho)$.

A natural question arises whether the statistics generated by an instrument assemblage can be reproduced by a single instrument $\{\mathcal{G}_{\lambda}\}_{\lambda}$ followed by classical post-processing.
Concretely, $\boldsymbol{\Lambda}$ is compatible if
\begin{equation}
    \Lambda_{a|x}=\sum_{\lambda} p(a|x,\lambda)\mathcal{G}_{\lambda}\quad \forall a,x,\label{eq: joint instrument}
\end{equation}
and incompatible otherwise (c.f. the channel-assemblage definition and unsteerability criterion given in Ref.~\cite{Piani2015}).
By applying Result 1, we label the instrument incompatibility witness $\mathcal{I}_g(\boldsymbol{\Lambda})$, as shown in Eq.~\eqref{eq: violation}.

To demonstrate the violation degree of instrument incompatibility, let us consider the channel steering scenario~\cite{Piani2015}: Charlie (C) prepares an input $\rho^{\rm C}$ and sends it to Bob by a fixed total channel $\Lambda^{\rm C\to B}$ which might be correlated with an ancillary system of, say, Alice.
The tasks is to decide if the correlation between Alice and Bob allows Alice to steer Bob's outputs of the channel.
In such a scenario, the instrument assemblage can be defined through an extended channel $\Lambda^{\rm C\to A\otimes B}$, reads
\begin{equation}
    \Lambda_{a|x}(\rho^{\rm C}) = \Tr_{\rm A}\, (M_{a|x}\otimes \id )\, \Lambda^{\rm C\to A\otimes B}(\rho^{\rm C}),\label{eq: channel steering}
\end{equation}
with $\sum_{a}\Lambda_{a|x}=\Lambda^{\rm C\to B}$.
Following Ref.~\cite{Uola2018pra}, every channel $\Lambda^{\rm C\to B}$ admits a Stinespring isometry $V:\mathcal H^{\rm C} \to  \mathcal H^{\rm A} \otimes \mathcal H^{\rm B}$. 
A minimal dilation $V^{\min}$ is obtained by choosing the ancilla dimension equal to the Kraus (Choi) rank $r$ of $\Lambda^{\rm C\to B}$, namely
$V^{\min}\ket{\psi}=\sum_{i=0}^{r-1}\ket{i} \otimes K_i\ket{\psi}$ with a linearly independent set of Kraus operators $\{K_i\}_{i=0}^{r-1}$ and $\sum_i K_i^\dagger K_i=\mathbb I$.
In this setting, any instrument assemblage $\boldsymbol{\Lambda}$ on C to B can be represented via a measurement assemblage $\boldsymbol{M}$ on the ancilla (A) as in Eq.~\eqref{eq: channel steering}.
The minimal dilation ensures a compatibility equivalence: $\boldsymbol{\Lambda}$ is unsteerable (i.e., compatible) if and only if the associated $\boldsymbol{M}$ is compatible. 
Consequently, instrument-level incompatibility can be certified by optimizing any measurement-level witness over all minimal dilations. 
Writing $\boldsymbol{M}(\boldsymbol{\Lambda},V^{\min},\rho^{\rm C})$ for the measurement assemblage associated with $\boldsymbol{\Lambda}$ through $V^{\min}$ and $\rho$, we define
\begin{equation}
\mathcal I_g(\boldsymbol{\Lambda})
  := \sup_{\rho^{\rm C}, \boldsymbol{M}} 
  \mathcal M_g \left[\boldsymbol{M}(\boldsymbol{\Lambda},V^{\min},\rho^{\rm C})\right],
\end{equation}
which is vanishing for compatible instruments, convex, and post-processing monotonic.
By optimizing the measurement assemblage $\boldsymbol{M}$, the correlations between A and B certify instrument incompatibility--and, whenever such a violation occurs, also certify that $\Lambda^{\rm C\to B}$ is not incompatibility-breaking~\cite{Ku2022PRXQ}.


As a illustrative example, we consider the total channel be a qubit-amplitude-damping channel $\Lambda^{\rm C\to B}\to \Lambda^{\rm amp}_\gamma$ with visibility $\gamma$, and we set the input state $\rho^{\rm C} \to \ket{1}$.
In such an example, $V^{\min}_\gamma \ket{1}=\sum_{i=0}^{1}\ket{i}\otimes \sqrt{\alpha_i}\,\ket{i\oplus 1}$, where $\alpha_0 = 1-\gamma$ and $\alpha_1 = \gamma$.
Together with the set of two noisy qubit Pauli measurements defined in Eq.~\eqref{eq: noisy Pauli} (see the Appendix~\ref{appendix: instrument assemblage} for details).
The results of $\mathcal{M}_{I}(\gamma, w)$ and $\mathcal{M}_{N}(\gamma, w)$ are shown in Figs.~\ref{fig2}\red{(c)} and \ref{fig2}\red{(d)}, respectively.
In both figures, the violation degrees disappear when $\gamma = \{0,1\}$ due to the minimal dilation of the channel $\Lambda^{\rm amp}_{\gamma=0,1}$ is a single unitary, i.e., one-dimensional, such that $V^{\min}\ket{\psi}=\ket{0}\otimes U\ket{\psi}$ is separable.
When $w=0$ (the red curves shown in each figures), both $\mathcal M_I|_{w=0}$ and $\mathcal M_N|_{w=0}$ reduce to instrument incompatibility parameters $\mathcal{I}_I(\gamma)$ and $\mathcal{I}_N(\gamma)$, indicating the correlation between Alice and Bob generated by the extended channel $\Lambda^{\rm C\to A\otimes B}$.
Note that, when $\gamma = 0.5$, the minimal dilation maps the input state $\ket{1}$ to maximally entangled state,  representing measurement incompatibility $\mathcal M_g(\boldsymbol{M})$.

\textit{Conclusion.}---We introduce a framework based on quantum operational information tasks to certify incompatibility and, from it, an optimization-free recipe: any convex functional $g$ on the relevant object space induces a compatibility bound, whose violation signals quantum incompatibility.
We establish a necessary and sufficient criterion for non-triviality: witnesses built from $g$ are nontrivial iff $g$ is non-affine on extremal points, underscoring the generality of the convex-analytic approach. 
As concrete examples, we present functionals based on the Wigner–Yanase skew information and $\ell_{2}$-type coherence that efficiently certify incompatibility in a dimension-independent manner.

For pure bipartite states, any violation lower-bounds the corresponding entanglement measure, and achievability of the funtional's roof extension ensures faithfulness. Characterizing sufficient conditions for quantifying entanglement of general mixed states, however, remains an open problem, which we leave as a future work.

Beyond states, the same recipe defines a steering-induced incompatibility monotone for measurements and instruments, and any observed violation can be read as preservation of incompatibility--in particular, a certificate that the channel is \textit{not} incompatibility-breaking~\cite{Ku2022PRXQ}.


\textit{Acknowledgments.}---K.-Y.L. acknowledges financial support from the European Union (ERC StG ETQO, Grant Agreement no.\ 101165230).
This work is supported by the National Center for Theoretical Sciences and National Science and Technology Council, Taiwan, Grant: NSTC 114-2112-M-006 -015 -MY3.
A.M. was supported by the Polish National Science Centre (NCN) under the Maestro Grant No. DEC-2019/34/A/ST2/00081.


%

\begin{appendix}
\begin{widetext}
    
\section{Derivation of a class of incompatibility witnesses}\label{appendix: HO ineq}
Hereafter we derive the incompatibility witness.
Given that the assemblage $\boldsymbol{O}$ admits the HO model, we obtain
\begin{equation}
\begin{aligned}
    g^{\rm as}(\boldsymbol{O}_{x}) 
    &= \sum_{a} p(a|x) g \left[ \sum_{\lambda}p(\lambda|a,x)O_{\lambda} \right] \\
    &= \sum_{a} p(a|x) g \left[ \sum_{\lambda}\frac{p(a|x,\lambda)p(\lambda)}{p(a|x)}O_{\lambda} \right] \\
    &\leq \sum_{a} p(a|x) \sum_{\lambda}\frac{p(a|x,\lambda)p(\lambda)}{p(a|x)} g(O_{\lambda}) \\
    &= \sum_{\lambda} p(\lambda) g(O_{\lambda}) \\
    &\leq \sum_{\lambda} p(\lambda) F_{g}(O_{\lambda}),
\end{aligned}
\end{equation}
where the first inequality follows from the convexity of $g$ and the second from the definition $F_{g}\ge g$.
Because $F_{g}$ is concave in $\mathcal{O}$, we also have
\begin{equation}
    \sum_{\lambda} p(\lambda)\,F_{g}(O_{\lambda})
    \le F_{g}^{\rm as}(\boldsymbol{O}_{x}) \quad \forall x.
\end{equation}
Combining the two chains of inequalities yields, for every assemblage $\boldsymbol{O}$ that admits the HO model,
\begin{equation}
    g^{\rm as}(\boldsymbol{O}_{x}) \stackrel{\text{\tiny{HO}}}{\leq} F_{g}^{\rm as}(\boldsymbol{O}_{x'}) \quad \forall x,x'.
\end{equation}
Hence the pair $(g^{\rm as},F_{g}^{\rm as})$ constitutes a valid compatibility witness.  
Analogously, starting from a concave functional $f$ and its (least) convex roof $G_{f}\ge f$, one obtains another class of incompatibility witness
\begin{equation}
    f^{\rm as}(\boldsymbol{O}_{x})
    \stackrel{\text{\tiny{HO}}}{\ge}
    G_{f}^{\rm as}(\boldsymbol{O}_{x'}) \quad \forall\,x,x'.
\end{equation}
Below we focus on the convex-$g$ version; the concave-$f$ case can be derived analogously.
As shown in the main text, the incompatibility witness can be defined by
\begin{equation}
    \mathcal{V}_{g}(\boldsymbol{O}) := \max \{ \max_{x} g^{\rm as}(\boldsymbol{O}_{x}) - \min_{x} F_{g}^{\rm as}(\boldsymbol{O}_{x}), 0 \}.
\end{equation}

For completeness, in the following, we detail the process of constructing a witness in a quantum incompatibility task and demonstrate that the maximizations and minimizations in the method are computationally efficient.

Consider a quantum device with $m$ classical inputs (labeled $x \in \{1, 2, \dots, m\}$) and $n$ classical outputs (labeled $a \in \{1, 2, \dots, n\}$). 
For each chosen input $x$, the device stochastically produces an output $a$ and a conditional quantum state $\rho_{a|x}$. 
This experimental setup provides a state assemblage $\boldsymbol{\sigma} = \{ \sigma_{a|x} = p(a|x) \rho_{a|x} \}_{a,x}$, where all $x$ settings share the same marginal state, i.e.,
\begin{equation}
    \sum_{a}\sigma_{a|x}=\sum_{a'}\sigma_{a'|x'} \quad \forall\, x,x' \in \{1,2,\dots,m\},
\end{equation}
noting that if the marginal states differ, the assemblage is trivially incompatible.

For each setting $x$, one calculates the assemblage versions of the functionals:
$g^{\rm as}(\boldsymbol{\sigma}_x)$ and $F_g^{\rm as}(\boldsymbol{\sigma}_x)$.
The violation degree of the incompatibility witness $\mathcal{S}_g(\boldsymbol{\sigma})$ is then simply given by:
\begin{equation}
\begin{aligned}
    \mathcal{S}_g(\boldsymbol{\sigma})
    &:=\max\left\{ \max_x g^{\rm as}(\boldsymbol{\sigma}_{x}) - \min_{x'}F_g^{\rm as}(\boldsymbol{\sigma}_{x'}) \, ,0\right\}  \\
    &=\max \left\{ \max \left\{g^{\rm as}(\boldsymbol{\sigma}_{1}),\dots, g^{\rm as}(\boldsymbol{\sigma}_{m}) \right\}-\min\left\{ F_g^{\rm as}(\boldsymbol{\sigma}_{1}),\dots,F_g^{\rm as}(\boldsymbol{\sigma}_{m})  \right\},\,0\right\}.
\end{aligned}
\end{equation}
As shown above, the required maximization and minimization involve only simple comparisons among $m$ real numbers.
The computational complexity scales linearly with the number of measurement settings $m$, and the associated computational tasks are negligible compared to existing optimization-based certification techniques~\cite{Cavalcanti_2016}.

\section{Proof of Result 2: The Nontriviality of Witnesses}\label{appendix: nontrivial}
We prove Result 2 by the following equivalent statement: $\mathcal{V}_{g}(\boldsymbol{O})=0 ~~\forall\boldsymbol{O}$ if and only if there exists an affine functional $L$ such that $g(O_{\rm ext})=L(O_{\rm ext})$. 
Moreover, let $\psi\in O_{\rm ext}$.
\begin{proof}---We start by analyzing the \textit{if} direction, we have $g(O_{\rm ext})=L(O_{\rm ext})=F_{g}(O_{\rm ext})$, such that $L(\sum_i p_i O_i)=\sum_i p_i L(O_i)$ for all decompositions.
Then
\begin{equation}
\begin{aligned}
    \max_{\boldsymbol{O}} \mathcal{V}_{g}(\boldsymbol{O}) 
    &=\max_{\{p(a|x),\psi_{a|x}\}_{a,x}} \left\{ \max_{x} \sum_{a} p(a|x) g(O_{a|x}) - \min_{x} \sum_{a} p(a|x)F_{g}(O_{a|x}) ,0 \right\} \\
    &\leq  \max\left\{\max_{\{p(a|x),\psi_{a|x}\}_{a,x}} \sum_{a} p(a|x) g(\psi_{a|x}) - \min_{\{p(a|x),\psi_{a|x}\}_{a,x}} \sum_{a} p(a|x)L(\psi_{a|x}) ,0\right\} \\
    &=  \max_{\{p(a|x),\psi_{a|x}\}_{a,x}} \sum_{a} p(a|x) L(\psi_{a|x}) - \min_{\{p(a|x),\psi_{a|x}\}_{a,x}} \sum_{a} p(a|x)L(\psi_{a|x})  \\
    &=  \max_{\{p(a|x),\psi_{a|x}\}_{a}} L\left[\sum_{a} p(a|x)\psi_{a|x}\right] - \min_{\{p(a|x),\psi_{a|x}\}_{a}} L\left[\sum_{a} p(a|x)\psi_{a|x}\right] \\
    &=  \max_{\{p(a|x),\psi_{a|x}\}_{a}}L(O^{\rm m}) - \min_{\{p(a|x),\psi_{a|x}\}_{a}} L(O^{\rm m}) \\
    &= 0,
\end{aligned}
\end{equation}
where, in the first line, the maximum taken over all assemblages $\boldsymbol{O}$ becomes the best decomposition of $\{p(a|x),\psi_{a|x}\}$ and taking the positive part; the second line comes from the property of the max operation, i.e., $\max\{X-Y\}\le \max X - \min Y$; the third line results from the non-negativity of the expression; the fifth line is given by the same marginal $O^{\rm m}$ for all $x$ [no-signaling condition: $\sum_{a} p(a|x)\psi_{a|x}=O^{\rm m}~\forall x$].
We note that in this proof, we exclude the case that $g$ being affine in the whole space $\mathcal{O}$, which directly makes the inequality trivial.

For the ``only if'' direction, assume $\mathcal V_{g}(\boldsymbol{O})=0$ for every assemblage $\boldsymbol{O}$ and fix the assemblage $\boldsymbol{O}^{\star}$ that maximizes the quantity $\max_{x}g^{\rm as}(\boldsymbol{O}_{x})-\min_{x'}F_{g}^{\rm as}(\boldsymbol{O}_{x'})$, which implies
\begin{equation}
    \max_{x} g^{\rm as}(\boldsymbol{O}^{\star}_{x})
    \leq
    \min_{x} F_{g}^{\rm as}(\boldsymbol{O}^{\star}_{x}).
\end{equation}
Given that $g^{\rm as}(\boldsymbol{O}^{\star}_{x})$ and 
$F_{g}^{\rm as}(\boldsymbol{O}^{\star}_{x})$ are the averages of
$g(\psi_{a|x})$ and $F_{g}(\psi_{a|x})$, respectively, over $\{\psi_{a|x}\}_a \subset \text{ext}(\mathcal{O})$ that decompose the same margin
$O^{\rm m}=\sum_a p(a|x)\psi_{a|x}~\forall x$. Thus, the above inequality implies
\begin{equation}
\begin{aligned}
    \max_{\{p(a|x),\psi_{a|x}\}_{a}}{\sum_a p(a|x) g(\psi_{a|x})}
    &\leq \min_{\{p(a|x),\psi_{a|x}\}_{a}}{\sum_a p(a|x) F_{g}(\psi_{a|x})}\\
    &= \min_{\{p(a|x),\psi_{a|x}\}_{a}}{\sum_a p(a|x) g(\psi_{a|x})},
\end{aligned}
\end{equation}
where the second line is given by $F_{g}(O_{\rm ext})=g(O_{\rm ext})$.
One can observe that the global maximum can not be smaller than the global minimum only if every decomposition of every admissible $O^{\rm m}$ yields the same value, i.e., the convex roof $g^{\cup}$ and the concave roof $g^{\cap}$ of $g$ coincide.
This suggests that the function $g$ is affine on $\text{ext}(\mathcal{O})$, which concludes the proof.

\end{proof}

\section{Examples of non-trivial inequalities for state assemblages}\label{appendix: examples}
Here, we provide two non-trivial examples of constructing incompatibility witnesses by a convex functional $g$.
We start by recalling the definition of the Wigner-Yanase skew information~\cite{Wigner1963,Luo2003prl}:
\begin{equation}
\begin{aligned}
    I(\rho,H)
    :=& - \frac{1}{2}\Tr[\sqrt{\rho},H]^{2} \\
    =&  \Tr \rho H^2 - \Tr \sqrt{\rho}H\sqrt{\rho}H,
\end{aligned}
\end{equation}
which is convex for states~\cite{Takagi2019sr}.
Then, we have
\begin{equation}
\begin{aligned}
    F_{I}(\rho,H)
    :=& \sup \sum_{i} p_i I(\psi_i ,H)\\
    =& \sup \sum_{i} p_i \left[ \bra{\psi_i}H^2 \ket{\psi_i} - \bra{\psi_i}H \ket{\psi_i}\bra{\psi_i}H \ket{\psi_i} \right] \\
    =& \sup \sum_{i} p_i \text{Var}(\psi_i ,H) \\
    =& \text{Var}(\rho ,H),
\end{aligned}
\end{equation}
where $\text{Var}(\rho ,H)$ is the variance of observable $H$, i.e., $\text{Var}(\rho,H)=\Tr(\rho H^2)-(\Tr\rho H)^2$, and the forth line is due to variance is its own concave roof~\cite{Toth2013pra}.
Therefore, the violation of $I$ can be given by
\begin{equation}
    \mathcal{S}_I(\boldsymbol{\sigma})=\max \{\max_x I^{\rm as}(\boldsymbol{\sigma}_x) - \min_x {\rm Var}^{\rm as}(\boldsymbol{\sigma}_x),0 \}.\label{eq: skew witness}
\end{equation}

The second illustrative example is the squared $\ell_2$-type of coherence for a given basis $\{\ket{i}\}_i$, which is defined as~\cite{Baumgratz2014PRL}
\begin{equation}
\begin{aligned}
    N(\rho,\boldsymbol{\Pi})
    :=& \sum_{i\neq j} |\bra{i} \rho \ket{j}|^2\\
    =& \sum_{i,j} \bra{i}\rho\ket{j}\bra{j}\rho\ket{i}-\sum_i \bra{i}\rho\ket{i}\bra{i}\rho\ket{i} \\
    =& \sum_i \Tr~\rho^2\Pi_i-\Tr~\rho\Pi_i\rho\Pi_i, \label{eq: l2 norm}
\end{aligned}
\end{equation}
where $\boldsymbol{\Pi}=\{\Pi_i = \ket{i}\bra{i}\}_i$ and its concave extension reads
\begin{equation}
\begin{aligned}
    F_N(\rho,\boldsymbol{\Pi})
    :=& \sup \sum_i p_i N(\psi_i,\boldsymbol{\Pi})\\
    =& \sup \sum_i p_i \left(\sum_j\bra{\psi_i}\Pi_j\ket{\psi_i} - \bra{\psi_i}\Pi_j\ket{\psi_i}\bra{\psi_i}\Pi_j\ket{\psi_i}\right)\\
    =& \sup \sum_i p_i \sum_j{\rm Var}(\psi_i , \Pi_j)\\
    =& \sum_j{\rm Var}(\rho , \Pi_j)\\
    :=& \overline{\rm Var}(\rho,\boldsymbol{\Pi}). \label{eq: F_N}
\end{aligned}
\end{equation}
The corresponding violation degree reads
\begin{equation}
    \mathcal{S}_N(\boldsymbol{\sigma})=\max \{\max_x N^{\rm as}(\boldsymbol{\sigma}_x) - \min_x \overline{\rm Var}^{\rm as}(\boldsymbol{\sigma}_x),0 \}.
\end{equation}

Additionally, we wish to emphasize that these functionals, i.e., the WYSI and $\ell_2$-type coherence, are experimental accessible, as they can be estimated without requiring full quantum state tomography.
As shown in Ref.~\cite{Girolami2014prl}, the lower bound of WYSI, $I^{L}(\rho,H)$, is directly measurable: 
\begin{equation}
    I^{L}(\rho,H) = -\frac{1}{4}\Tr[\rho,H]^2\le I(\rho,H).
\end{equation}
This indicates we can use this lower bound to construct a measurable witness that is sufficient to demonstration the violation of $\mathcal{S}_I$ in Eq.~\eqref{eq: skew witness}.

Furthermore, we note that this lower bound $I^{L}(\rho,H)$ is directly related to the $\ell_{2}$-type coherence $N(\rho,\boldsymbol{\Pi})$ by replacing the observable $H$ with a set of projective measurements $\boldsymbol{\Pi}=\{\Pi_{i}\}_{i}$. Specifically:
\begin{equation}
\begin{aligned}
    N(\rho,\boldsymbol{\Pi}) 
    &= \sum_{i \neq j} |\bra{i}\rho\ket{j}|^2 \\
    &= \sum_i \Tr \,\rho^2 \Pi_i - \Tr\, \rho \Pi_i \rho\Pi_i \\
    &= \sum_i 2\, I^{L}(\rho,\Pi_i).
\end{aligned}
\end{equation}
Consequently, both witnesses are physically measurable, rendering them highly practical for experimental implementation.

\section{Proof of Result 3: The Witnesses Lower bounds a full entanglement measure}\label{appendix: entanglement measure}

Here, we consider the case that Alice and Bob share a pure entangled state $\rho^{\rm AB}$.
We first show that to obtain the maximal violation degree, the optimal measurements Alice performed are set at most rank one, i.e., ${\rm rank}(M_{a|x})=1~\forall a,x$.
Let's recall that $\sigma_{a|x}=\Tr_{\rm A}[(M_{a|x}\otimes \id)\, \rho^{\rm AB}]$:
\begin{equation}
\begin{aligned}
    g^{\rm as}(\boldsymbol{\sigma}_{x}) 
    &:= \sum_{a}p(a|x)g(\rho_{a|x}) \\
    &= \sum_{a} p(a|x) g\left[ \frac{{\rm Tr}_{A}(M_{a|x}\ts \id) \rho^{\rm AB}}{p(a|x)} \right],
\end{aligned}
\end{equation}
where $p(a|x)={\rm Tr}\left[(M_{a|x}\ts \id) \rho^{\rm AB} \right]$.
Here, we can decompose each measurement $M_{a|x}$ into fine-grained measurement $M_{i,a|x}\ge 0$ and ${\rm rng}(M_{i,a|x})\le 1~\forall i,a,x$ such that $\sum_{i} M_{i,a|x} = M_{a|x}~\forall a,x$.
Under this assumption, we have a fine-grained version of $g^{\rm as}$, that is,
\begin{equation}
\begin{aligned}
    g_{\rm fine}^{\rm as}(\boldsymbol{\sigma}_{x})
    &=\sum_{i,a} p(i,a|x) g\left[ \frac{{\rm Tr}_{A}(M_{i,a|x}\ts \id) \rho^{\rm AB}}{p(i,a|x)} \right] \\
    &= \sum_{a} p(a|x) \sum_{i}\frac{p(i,a|x)}{p(a|x)}g\left[ \frac{{\rm Tr}_{A}(M_{i,a|x}\ts \id) \rho^{\rm AB}}{p(i,a|x)} \right] \\
    &\geq \sum_{a} p(a|x) g\left[\sum_{i}\frac{p(i,a|x)}{p(a|x)} \frac{{\rm Tr}_{A}(M_{i,a|x}\ts \id) \rho^{\rm AB}}{p(i,a|x)} \right] \\
    &= \sum_{a} p(a|x) g\left[ \frac{\sum_{i}{\rm Tr}_{A}(M_{i,a|x}\ts \id) \rho^{\rm AB}}{p(a|x)} \right] \\
    &= g^{\rm as}(\boldsymbol{\sigma}_{x}),
\end{aligned}
\end{equation}
where $p(i,a|x)={\rm Tr}\left[(M_{i,a|x}\ts \id) \rho^{\rm AB} \right]$.
The above inequality is obtained by the convexity; therefore, one can readily obtain $F^{\rm as}_{g,\textrm{fine}}\le F^{\rm as}_{g}$ by its concavity and conclude that the optimal measurement should be rank one.
Thus, in the following, we let ${\rm rank}(M_{a|x})=1~\forall a,x$.

Combining with the case that $\rho^{\rm AB}\to \psi^{AB}$ is pure,
we have the conditional states $\{\psi_{a|x}\}_{a,x}$ on Bob's side are pure.
In such case, we can further compute the optimal violation degree of pure bipartite input $\psi^{AB}$ as follows:
\begin{equation}
\begin{aligned}
    \mathcal{S}_{g}(\psi^{AB}) 
    &= \max \left\{  \max_{\{p(a|x),\psi_{a|x}\}} \sum_{a} p(a|x) g(\psi_{a|x}) - \min_{\{p(a|x),\psi_{a|x}\}} \sum_{a}p(a|x)F_{g}(\psi_{a|x})  ,0 \right\} \\
    &=   \max_{\{p(a|x),\psi_{a|x}\}} \sum_{a} p(a|x) F_g(\psi_{a|x}) - \min_{\{p(a|x),\psi_{a|x}\}} \sum_{a}p(a|x)g(\psi_{a|x})  \\
    &\le F_{g}\left[\sum_{a} p(a|x^{\star}) \psi_{a|x^{\star}}\right] - g\left[\sum_{a}p(a|x_{\star})\psi_{a|x_{\star}}\right]\\
    &= F_{g} (\rho^{\rm B}) - g(\rho^{\rm B}),
\end{aligned}
\end{equation}
where the second line is given by $F(\psi)=g(\psi)$ such that it is non-negative; the third line is taken by optimized when $x=x^{\star}$ ($x=x_{\star}$) is maximum (minimum) function, respectively, and the equation holds if the concave (convex) roof is achievable.
According to Ref.~\cite{Horodecki2009rmp}, a function of $\rho^{\rm B}$ satisfies strong monotonicity under local operation an classical communication (LOCC) if and only if (i) it is concave, and (ii) it is symmetry (under permutations of its arguments) and expandable.
For (i), notice that given $g$ is convex and $F_g$ is concave, therefore
\begin{equation}
\begin{aligned}
    [F_{g}-g][t\rho_{1}+(1-t)\rho_{2}] 
    &= F_{g}[t\rho_{1}+(1-t)\rho_{2}] - g[t\rho_{1}+(1-t)\rho_{2}] \\
    &\geq tF_{g}(\rho_{1})+(1-t)F_{g}(\rho_{2}) - tg(\rho_{1}) - (1-t)g(\rho_{2}) \\
    &= t[F_{g}-g](\rho_{1})+(1-t)[F_{g}-g](\rho_{2}),
\end{aligned}
\end{equation}
which indicates $[F_{g}-g]$ is concave.
For (ii), as we require the function $g$ to be symmetric and expandable, its concave roof and the derivative function $[F_{g}-g]$ are also symmetric and expandable.
Thus, we can conclude that $\mathcal{S}_{g}$ is monotonic under LOCC for an given convex, symmetric, and expandable $g$.

In addition, any function that is convex and symmetric is Schur-convex, while any function that is concave and symmetric is Schur-concave~\cite{RobertsVarberg1973,Bhatia1997,Nielsen1999}. 
In our case, the quantity $[F_g-g]$ is concave and symmetric, hence it is Schur-concave in the spectrum of the relevant state (e.g., in the eigenvalues of $\rho_B$).
Since the maximally mixed spectrum $u=(1/d,\ldots,1/d)$ of a $d$-dimension system is majorized by every other spectrum ($u\prec p$), Schur-concavity implies $[F_g-g](u)\ge [F_g-g](p)$ for all $p$. 
Therefore the violation attains its maximum when the input marginal is maximally mixed. 
For pure bipartite inputs, $\rho_B$ is maximally mixed if and only if the global state is maximally entangled; consequently the algebraic maximum of the violation is achieved only for maximally entangled inputs, i.e.,
\begin{equation}
    \mathcal{S}_g^{\rm max} = [F_g - g](\frac{\id}{d}),
\end{equation}
with system dimension $d$.

\section{Proof of Result 4: A valid incompatibility monotone}\label{appendix: measurement assemblage}

We first show that when the shared state $\ket{\psi^{AB}}=\sum_i \sqrt{p_i} \ket{i}\ket{i}$ is full-ranked such that $\rho^{\rm B}= \Tr_{\rm A} \ket{\psi^{AB}}\bra{\psi^{AB}} = \sum_i p_i \ket{i}\bra{i}$, the SEO $\boldsymbol{B}=\{B_{a|x}\}_{a,x}$ is equivalent to Alice's measurement $\boldsymbol{M}=\{M_{a|x}\}_{a,x}$ up to its transpose, which is
\begin{equation}
\begin{aligned}
    B_{a|x} 
    &= (\rho^{\rm B})^{-1/2} \,\sigma_{a|x}(\rho^{\rm B})^{-1/2} \\
    &= \sum_m p_m^{-1/2} \ket{m}\bra{m}\left[ \text{Tr}_{A} ~(M_{a|x}\otimes \id )\sum_{i,j}\sqrt{p_i p_j} \ket{i}\bra{j} \otimes \ket{i}\bra{j} \right] \sum_n p_n^{-1/2} \ket{n}\bra{n}\\
    &=   \sum_{i,j,m,n} \sqrt{ q_i q_j} \, p_m^{-1/2}p_n^{-1/2}\bra{j} M_{a|x} \ket{i} \ket{m} \braket{m|i}\braket{j|n}\bra{n} \\
    &= \sum_{m,n} \bra{m}M_{a|x}^{T}\ket{n}\ket{m}\bra{n} \\
    &=  M_{a|x}^{T},
\end{aligned}
\end{equation}
where the second line holds due to $\ket{\psi^{AB}}$ is full-ranked, such that $p_i >0 ~~\forall i$.
In the following, thus, we replace the SEO input $\boldsymbol{B}$ by the Alice measurement assemblage $\boldsymbol{M}$.

According to Ref.~\cite{Paul2019prl}, $\mathcal{M}_{g}(\boldsymbol{M})$ is a valid incompatibility monotone if it satisfies:
\begin{itemize}
    \item[(a)] $\mathcal{M}_{g}(\boldsymbol{M})=0$, if $\boldsymbol{M}$ is jointly measurable.
    \item[(b)] $\mathcal{M}_{g}(\boldsymbol{M})$ satisfies convexity.
    \item[(c)] $\mathcal{M}_{g}(\boldsymbol{M})$ is non-increasing under post-processing, namely
    \begin{equation}
    M_{a'|x'}= 
    \mathcal{W}(M_{a|x}) = \sum_{a,x}p(x|x')p(a'|a,x,x')M_{a|x}\quad \forall a,x,\label{eq: wiring map for measurements}
    \end{equation}
\end{itemize}
where $ \mathcal{W}$ denotes post-processing with the conditional probabilities $p(x|x')$ and $p(a'|a,x,x')$~\cite{Gallego2015prx}.

The condition (a) is automatically satisfied by the definition of $\mathcal{M}_{g}$:
\begin{equation}
    \mathcal{M}_{g}(\boldsymbol{M}) = \sup_{\rho_B} \mathcal{S}_{g}[\sqrt{\rho_B}\,\boldsymbol{M} \sqrt{\rho_B}].
\end{equation}
Given that a measurement assemblage $\boldsymbol{M}$ is compatible if and only if its SEO induced a state assemblage $\boldsymbol{\sigma} = \sqrt{\rho_B} \,\boldsymbol{M} \sqrt{\rho_B}$ that can be described by an LHS model.
Thus, we conclude that $\mathcal{M}_{g}(\boldsymbol{M})=0$ if $\boldsymbol{M}$ is compatible.

For the condition (b) $\mathcal{M}_{g}(\boldsymbol{M})$ satisfies convexity, we only need to demonstrate $\mathcal{S}_{g}({\boldsymbol{\sigma}})$ is convex in assemblage.
To show this, let us consider a state assemblage
\begin{equation}
\boldsymbol{\sigma}
\to q\boldsymbol{\sigma}+(1-q)\boldsymbol{\sigma}^{\prime}
:=\{q\sigma_{a|x}+(1-q)\sigma_{a|x}^{\prime}\}_{a,x}\quad \forall q \in [0,1].    
\end{equation}
Given $g$ is convex, its assemblage version $g^{\rm as}$ is naturally convex, i.e.,
\begin{equation}
\begin{aligned}
    g^{\rm as}(\boldsymbol{\sigma}) 
    &\to g^{\rm as}\left[ q\boldsymbol{\sigma} + (1-q)\boldsymbol{\sigma}' \right] \\
    &= \max_{x}   \sum_a p(a|x) g\left[ q\rho_{a|x} + (1-q)\rho'_{a|x}\right] \\
    &\leq \max_{x} \sum_a \left[ q p(a|x)g(\rho_{a|x}) + (1-q)p'(a|x)g(\rho'_{a|x}) \right] \\ 
    &\leq q \max_{x}  \sum_a p(a|x)g(\rho_{a|x}) 
     + (1-q) \max_{x} \sum_a p'(a|x)g(\rho'_{a|x}) \\
    &= q g^{\rm as}(\boldsymbol{\sigma}) + (1-q) g^{\rm as}(\boldsymbol{\sigma}').
\end{aligned}
\end{equation}

Analogously, with the fact that the $F_{g}(\rho)$ is concave, we have that $F_{g}^{\rm as}(\boldsymbol{\sigma})$ is also concave with respect to a combination of the state assemblages $q\boldsymbol{\sigma}+(1-q)\boldsymbol{\sigma}^{\prime}$.

Then, we can show that the violation degree satisfies convexity, namely 
\begin{equation}
\begin{aligned}
    \mathcal{S}_{g}(\boldsymbol{\sigma})
    &\to \mathcal{S}_{g}\left[ q\boldsymbol{\sigma}+(1-q)\boldsymbol{\sigma}' \right] \\
    &= \max \left\{ g^{\rm as}[q\boldsymbol{\sigma}+(1-q)\boldsymbol{\sigma}'] - F_{g}^{\rm as}[q\boldsymbol{\sigma}+(1-q)\boldsymbol{\sigma}'],0 \right\} \\
    &\leq \max \left\{  q\left[ g^{\rm as}(\boldsymbol{\sigma}) - F_{g}^{\rm as}(\boldsymbol{\sigma}) \right] 
     + (1-q)\left[ g^{\rm as}(\boldsymbol{\sigma}') - F_{g}^{\rm as}(\boldsymbol{\sigma}') \right] ,0\right\} \\
    &\leq q\max \left\{  \ g^{\rm as}(\boldsymbol{\sigma}) - F_{g}^{\rm as} (\boldsymbol{\sigma}) ,0\right\} 
     + (1-q) \max \left\{  g^{\rm as}(\boldsymbol{\sigma}') - F_{g}^{\rm as} (\boldsymbol{\sigma}'),0 \right\} \\
    &=q\mathcal{S}_{g}(\boldsymbol{\sigma}) + (1-q)\mathcal{S}_{g}(\boldsymbol{\sigma}'),\label{eq: convexity}
\end{aligned}
\end{equation}
where the third line is given by $g$ ($F_{g}^{\rm as}$) is, respectively, convex (concave), and the fourth line is the property of the maximization.
Therefore, we conclude the proof that $\mathcal{M}_{g}(\boldsymbol{M})$ is a convex functional.

Finally, for condition (c), we consider a post-processing $\mathcal{W}$ acting on $\boldsymbol{M}$ can be written as
\begin{equation}
\begin{aligned}
    M_{a'|x'}^T
    &=\mathcal{W}(M_{a|x}^T) \\
    &= \sum_{a,x}p(x|x')p(a'|a,x,x')M_{a|x}^T \\
    &= \sum_{a,x}p(x|x')p(a'|a,x,x')(\rho^{\rm B})^{-1/2} \,\sigma_{a|x}(\rho^{\rm B})^{-1/2} \\
    &=(\rho^{\rm B})^{-1/2} \mathcal{W}(\sigma_{a|x})(\rho^{\rm B})^{-1/2},
    \label{eq: wiring map}
\end{aligned}
\end{equation}
in which we obtain
\begin{equation}
    \sqrt{\rho^{\rm B}} \,\mathcal{W}(\boldsymbol{M})\sqrt{\rho^{\rm B}} = \mathcal{W}(\boldsymbol{\sigma}).
\end{equation}
Due to the convexity of $\mathcal{S}_{g}$, this implies  $\mathcal{S}_{g}$ is also non-increasing under post-processing, i.e., $\mathcal{S}_{g}[\mathcal{W}(\boldsymbol{\sigma})]\le \mathcal{S}_{g}(\boldsymbol{\sigma})$.
By using the above results, we can therefore show that $\mathcal{M}_{g}(\boldsymbol{M})$ is also non-increasing under $\mathcal{W}$:
\begin{equation}
\begin{aligned}
    \mathcal{M}_{g}\left[\mathcal{W}(\boldsymbol{M})\right] 
    &= \sup_{\rho^{\rm B}} \mathcal{S}_{g} \left[ \sqrt{\rho^{\rm B}} \,\mathcal{W}(\boldsymbol{M})\sqrt{\rho^{\rm B}}\right] \\
    &= \mathcal{S}_{g} \left[ \sqrt{\rho^{B\star}}\, \mathcal{W}(\boldsymbol{M})\sqrt{\rho^{B\star}}\right] \\
    &= \mathcal{S}_{g} \left[ \mathcal{W}\left(\sqrt{\rho^{B\star}}\, \boldsymbol{M}\sqrt{\rho^{B\star}}\right)\right] \\
    &\leq \mathcal{S}_{g}\left(\sqrt{\rho^{B\star}}\, \boldsymbol{M}\sqrt{\rho^{B\star}}\right) \\
    &\leq \sup_{\rho^{\rm B}}\mathcal{S}_{g} \left(\sqrt{\rho^{\rm B}} \,\boldsymbol{M}\sqrt{\rho^{\rm B}}\right) \\
    &= \mathcal{M}_{g}(\boldsymbol{M}), 
\end{aligned}
\end{equation}
where $\rho^{B\star}$ is the optimal argument in the supremum.

\section{Example of qubit Pauli measurements for pure entangled state input}\label{appendix: measurement assemblage example}

Here we illustrate the construction using qubit Pauli measurements and a pure entangled input state $\ket{\varphi^{AB}(\theta)}=\sin\theta\ket{00} + \cos\theta\ket{11}$ with noisy Pauli measurements $M_{a|x}^w=(1-w)\Pi_{a|x}+w\,\id/2$. 
Taking the supremum over $\rho^{\rm B}$, we set $\theta=\pi/4$ and obtain
\begin{equation}
\begin{aligned}
    \sigma_{a|x} 
    &= \Tr_{\rm A} \left(M_{a|x}^w \otimes \id  \right)\frac{1}{2}\sum_{i,j}  \ket{i}\bra{j}\otimes  \ket{i}\bra{j} \\
    &=\frac{1}{2} \sum_{i,j} \bra{j}M_{a|x}^w \ket{i} \ket{i}\bra{j}\\
    &= \frac{1}{2} M_{a|x}^w,
\end{aligned}
\end{equation}
where in the third line we used the property that the (noisy) qubit Pauli effects are Hermitian, $M_{a|x}^{w\,\dagger}=M_{a|x}^w$.
In this example the conditional probabilities are uniform, $p(a|x)=\Tr M_{a|x}^w /2=1/2$ for all $a,x$, and $\rho_{a|x}(w)=M_{a|x}^w$ is a classical mixture of the projector $\Pi_{a|x}$ and the maximally mixed state $\id/2$, with $w\in[0,1]$.

For $\mathcal M_N(w)$, we obtain an analytical expression that matches the known joint-measurability threshold for two Pauli measurements, namely $w\le 1-1/\sqrt{2}$~\cite{Heinosaari2008}.
Starting from the definition of $N^{\rm as}(\boldsymbol{\sigma}_x,\boldsymbol{\Pi})$ in Eq.~\eqref{eq: l2 norm}, we obtain
\begin{equation}
\begin{aligned}
    N^{\rm as}(\boldsymbol{\sigma}_x,\boldsymbol{\Pi}) 
    &= \sum_{a} \frac{1}{2} N\left[\rho_{a|x}(w),\boldsymbol{\Pi}\right] \\
    &= \sum_a \frac{1}{2} \sum_i \left[ \bra{i}\rho_{a|x}(w)^2\ket{i} - \bra{i}\rho_{a|x}(w)\ket{i}^2   \right] \\
    &= \sum_{a} \frac{1}{2}\sum_i \left\{ \bra{i}\left[ (1-w)^2 \Pi_{a|x} + w(1-w)\Pi_{a|x} + w^2 \frac{\id}{4} \right] \ket{i} - \left[ \bra{i}(1-w)\Pi_{a|x}\ket{i}+\frac{w}{2} \right]^2   \right\} \\
    &= \sum_{a} \frac{1}{2}  \left\{(1-w)\sum_i\left[ \bra{i}  \Pi_{a|x} \ket{i}\right]  -(1-w)^2 \sum_i \left[ \bra{i}\Pi_{a|x}\ket{i}^2 \right] - w(1-w)\sum_i\left[ \bra{i}\Pi_{a|x}\ket{i} \right]   \right\} \\
    &= \sum_{a} \frac{1}{2}  \left\{(1-w) - w(1-w) -(1-w)^2 \sum_i \bra{i}\Pi_{a|x}\ket{i}^2 \right\} \\
    &= (1-w)  - w(1-w) -  \frac{(1-w)^2}{2}\sum_{a,i}  \bra{i}\Pi_{a|x}\ket{i}^2.
\end{aligned}
\end{equation}
Similarly, substituting $\rho_{a|x}(w)$ into the concave roof $F_N$ in Eq.~\eqref{eq: F_N} yields
\begin{equation}
\begin{aligned}
     \overbar{\rm Var}^{\rm as}(\boldsymbol{\sigma}_{x'},\boldsymbol{\Pi}) 
     &=  \sum_{a} \frac{1}{2} \overbar{\rm Var}\left[\rho_{a|x'}(w),\boldsymbol{\Pi}\right] \\
     &=  \sum_{a} \frac{1}{2} \sum_i \left\{ \bra{i}\rho_{a|x'}(w)\ket{i} - \left[ \bra{i}\rho_{a|x'}(w)\ket{i} \right]^2 \right\} \\
     &= \sum_{a} \frac{1}{2}\left\{1 - \sum_i \bra{i} \left[ (1-w) \Pi_{a|x'} + w\frac{\id}{2} ) \right]\ket{i}^2 \right\} \\
     &= \sum_{a} \frac{1}{2}\left\{1 -(1-w)^2 \sum_i\left[ \bra{i}  \Pi_{a|x'}\ket{i}^2 \right] - w(1-w)\sum_i\left[ \bra{i}  \Pi_{a|x'}\ket{i} \right] - \frac{w^2}{2} \right\} \\
     &= 1 - \frac{w^2}{2} - w(1-w) -\frac{(1-w)^2}{2} \sum_{a,i} \bra{i}  \Pi_{a|x'}\ket{i}^2.
\end{aligned}
\end{equation}
By the definition of the violation degree, we have
\begin{equation}
\begin{aligned}
    \mathcal{M}_{N}(w) 
    &= \max \left\{  \max_{x} N^{\rm as}(\boldsymbol{\sigma}_x,\boldsymbol{\Pi}) - \min_{x'} \overbar{\rm Var}^{\rm as}(\boldsymbol{\sigma}_{x'},\boldsymbol{\Pi}),0 \right\} \\
    &= \max \left\{ \frac{w^2}{2}-w - \frac{(1-w)^2}{2}\left( \min_x\sum_{a,i} \bra{i}  \Pi_{a|x}\ket{i}^2 - \max_{x'} \sum_{a,i} \bra{i}  \Pi_{a|x'}\ket{i}^2  \right),0\right\}.
\end{aligned}
\end{equation}
In the second line, the minimum (maximum) arises because the last term in $N^{\rm as}(\boldsymbol{\sigma}_x,\boldsymbol{\Pi})$ ($\overbar{\rm Var}^{\rm as}(\boldsymbol{\sigma}_{x'},\boldsymbol{\Pi})$) is negative, so the extremizers are exchanged. 
Since $\{\Pi_{a|x}\}_x$ are the eigenprojectors of the Pauli $Z$ and $X$ observables, the quantity $\sum_{a,i} \bra{i}  \Pi_{a|x}\ket{i}^2$ takes the values $\{2,1\}$.
Inserting these values in and imposing $\mathcal M_g(w)\ge 0$ gives
\begin{equation}
\begin{aligned}
     0 &\leq \frac{w^2}{2} - w + \frac{(1-w)^2}{2} \\
     &= w^2 -2w + \frac{1}{2},
\end{aligned}
\end{equation}
from which we find that $\mathcal M_g(w)\ge 0$ for $w \in (-\infty ,1 - 1/\sqrt{2}]$ and for $w \in [1 + 1/\sqrt{2}, \infty )$; restricting to $w\in[0,1]$ yields the threshold $w\le 1-1/\sqrt{2}$.

\section{Instrument assemblage and channel steering under amplitude damping channel and noisy Pauli measurements}\label{appendix: instrument assemblage}

Here, as a concrete example, we consider the total channel $\Lambda^{C \to B} \to \Lambda^{\rm amp}_{\gamma}$ to be a qubit-amplitude-damping channel with Kraus operators
\begin{equation}
    K_{0} = 
    \left(\begin{matrix}
        1 & 0 \\
        0 & \sqrt{1-\gamma}
    \end{matrix}\right)\quad\text{and}\quad
    K_{1}= 
    \left(\begin{matrix}
        0 & \sqrt{\gamma} \\
        0 & 0
    \end{matrix}\right)\quad \text{with} \quad \gamma \in [0,1].
\end{equation}
In such a case, the minimal dilation can be found as 
\begin{equation}
    V_\gamma \ket{\psi}= \sum_{i=0}^1 \ket{i}\otimes K_{i} \ket{\psi}.
\end{equation}
As we consider the input state $\rho^{\rm C} \to \ket{1}\bra{1}$, the extended channel $\Lambda^{C \to A \otimes B}$ can be written as 
\begin{equation}
\begin{aligned}
    \Lambda^{C \to A \otimes B}\left(\ket{1}\bra{1}\right) 
    &= V_\gamma \ket{1}\bra{1} V_\gamma^\dagger \\
    &= \sum_{i,j} \sqrt{\alpha_i \alpha_j}\,\ket{i}\bra{j}\otimes\ket{i\oplus1}\bra{j\oplus1}, 
\end{aligned}
\end{equation}
where $\alpha_0 = 1-\gamma$ and $\alpha_1 = \gamma$.
The output state from the extend channel is pure entangled, and  $\alpha_i$ represent the Schmidt coefficients.
For $\gamma \in \{0,1\}$, the output state is separable, and thus, indicating the impossibility for Alice to steer Bob's outputs of the channel.

Together with two noisy qubit Pauli measurements and inpute state $\ket{1}$, we can set the instrument assemblage as
\begin{equation}
\begin{aligned}
     \Lambda_{\gamma, a|x}^w (\ket{1}\bra{1})
     &= \Tr_{\rm A} \, (M^w_{a|x}\otimes \id ) V_\gamma \ket{1}\bra{1} V_\gamma^\dagger \\
     &= \Tr_{\rm A} \, (M^w_{a|x}\otimes \id ) \sum_{i,j} \ket{i}\bra{j}\otimes K_i \, \ket{1}\bra{1}\, K_j^{\dagger} \\
     &= \sum_{i,j} \bra{j}M_{a|x}^{w}\ket{i} \sqrt{\alpha_i \alpha_j}\ket{i\oplus1}\bra{j\oplus1},
\end{aligned}
\end{equation}
which satisfies $\sum_{a} \Lambda_{\gamma, a|x}^w = \Lambda^{\rm amp}_{\gamma}~~\forall x,w$.
We note that for $\gamma = 0.5$, $V_{\gamma = 0.5}$ directly maps $\ket{1}$ to a maximally entangled state, and consequently reproduces the result of the incompatible measurement assemblage as shown in Appendix.~\ref{appendix: measurement assemblage example}.

\end{widetext}
\end{appendix}

\end{document}